\documentclass[showpacs,aps,prb,reprint,floatfix]{revtex4-1}
\usepackage{graphicx}
\usepackage{amssymb}
\usepackage{amsmath}
\usepackage{dcolumn}
\usepackage[utf8]{inputenc}
\usepackage{bm}
\usepackage{color}
\usepackage{float}
\usepackage{physics}
\usepackage{tikz}
\usepackage{subfigure}

\begin{document}
\title{Magnetic phase separation in a frustrated ferrimagnetic chain under a magnetic field}

\author{A. M. do Nascimento-Junior}
\author{R. R. Montenegro-Filho}
\affiliation{Laborat\'{o}rio de F\'{i}sica Te\'{o}rica e Computacional, Departamento de F\'{i}sica, Universidade Federal de Pernambuco, 50760-901 Recife-PE, Brasil}
\date{\today}

\begin{abstract}
We use density matrix renormalization group to study the first-order quantum phase transition induced by a
magnetic field $h$ in a frustrated ferrimagnetic chain. The magnetization ($m$) curve as a function of $h$ presents
a macroscopic jump and the energy curve as a function of $m$ has two global minima. We characterize the two competing phases 
and study the phase-separated states in the coexistence region. Also, we observe that the transition is accompanied 
by an increase in the number of itinerant singlet pairs between sites in the unit cells of the chain. 
Finally, we identify the critical point at the end of the first-order transition line and a crossover line.
\end{abstract}

\maketitle
\section{Introduction} 
Systems made of interacting magnetic units, ions or atoms, can exhibit interesting collective
and local quantum modes as control parameters vary. The quantum transitions 
\cite{sachdev2001quantum,continentino2017quantum,Vojta2003} among these phases or states occur at zero temperature
 as doping, pressure or an applied magnetic field change. 
In a first-order transition \cite{Binder1987},
the first derivative of the energy as a function of the control parameter is discontinuous, giving rise to a discontinuous
change (a jump) in the conjugate variable, and interesting phenomena like phase separation and hysteresis. 
The physics of first-order quantum phase transitions is observed in 
strongly correlated electronic materials \cite{Dagotto2001,*Dagotto2005,Uemura2006,Macedo1995,*Montenegro-Filho2006,*Montenegro-Filho2014,Post2018} and 
atomic gases in optical lattices \cite{Partridge2006,Liao2011,Okumura2011,Landig2016,Campbell2016,Szirmai2017,Hruby2018,Wald2018}.   
In magnetic systems with exchange anisotropy, the metamagnetic transition \cite{Kohno1997,Sakai1999} 
is typically first order but can be continuous if quantum fluctuations are sufficiently strong \cite{Sakai1999}.

Geometrically frustrated magnetic systems \cite{Series2011,Balents2010} have  
exchange coupling patterns favoring the existence of competing ground states. 
Quantum phase transitions can occur as the couplings change in 
response to pressure variations, or by changing an applied magnetic field ($h$).
In a first-order phase transition with $h$, the magnetization $m$ as
a function of $h$ presents a macroscopic jump. 
In two dimensions, macroscopic magnetization jumps were
observed \cite{Takigawa2013,Matsuda2013} in the frustrated magnet SrCu$_2$(BO$_3$)$_2$, the orthogonal-dimer antiferromagnet, 
which is modeled by the Shastry-Sutherland lattice \cite{Wang2018},
and in kagom\'e lattices \cite{Picot2016,Nishimoto2013,Nakano2015,Hasegawa2018}. 
Recently it was evidenced that a ferrimagnetic ludwigite compound presents 
metamagnetic transitions \cite{Medrano2017} with phase coexistence due to magnetic ions in a frustrated state.

In one dimension, geometrically frustrated chains \cite{Chepiga2016,Derzhko2013,Montenegro-Filho2008,Pereira2009,
Torrico2016,Tenorio2009,Chandra2010,Furuya2014,JPSJ.86.084706} are more accessible to 
numerical or theoretical investigations and offer plenty of interesting physics. In zero magnetic field, first-order 
phase transitions induced by changes in the exchange parameters were identified in models of this type \cite{Kim2008,Montenegro-Filho2008,Pixley2014}. 
Under a magnetic field, macroscopic magnetization jumps
were observed in frustrated chains \cite{Schulenburg2002,Chandra2004}, including 
spin tubes \cite{GomezAlbarracin2014}, ladder models \cite{Honecker2000,PhysRevB.73.214405,Michaud2010,Elias2017} and 
systems with localized magnon states \cite{Richter2005}, as frustrated molecules \cite{Schroder2005}.

The phenomenon of phase separation, in which the two competing states
are observed in spatially separated regions, is a remarkable feature of first-order phase transitions. 
While magnetization jumps in the magnetization curves were observed in a 
variety of frustrated magnetic systems, the discussion of the phase-separated states of these quantum systems 
is less common. 

Here we use the density matrix renormalization group (DMRG) \cite{White1992,*White1993,*Schollwock2005,*Schollwock2011} to study the first-order transition induced by a magnetic field in a frustrated 
ferrimagnetic chain, particularly its phase-separated states. 
In Sec. II we present the Hamiltonian of the system and the details of the DMRG 
calculations. In Sec. III, we show the magnetization curve of the model and the magnetization jump for 
$J=0.7$, where $J$ is the frustration parameter. The associated energy curves 
for fixed values of the magnetic field, with the presence of metastable states, is also exhibited in this Section. 
In Sec. IV we characterize the two competing states in the transition by calculating average local magnetizations, transverse correlation 
functions, and local correlations. In particular, we show that the two competing states have a distinct density of itinerant singlet
pairs between two sites of a unit cell. 
Further, we present a detailed analysis of the states inside the jump and show that the two phases 
coexist in spatially separated regions of the chain. Also in
this Section, by changing $J$ we determine the critical point at which the jump closes. 
Finally, in Sec. V we summarize our results.

\section{Model and method} 
\begin{figure}[!htb]
\includegraphics[width=0.4\textwidth]{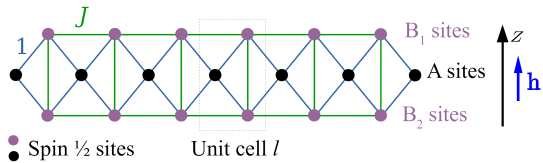}
\caption{Schematic representation of the Hamiltonian. The model has 
a $B$ sublattice with $B_1$ and $B_2$ sites in a ladder arrangement; and an $A$ sublattice, with 
a spin-$1/2$ in each site. The coupling 
between $A$ and $B$ spins defines the unit of energy; and 
$J$ is the coupling between $B$ spins. A magnetic field $\mathbf{h}$ is applied in the $z$-direction. 
In the density matrix renormalization group (DMRG) calculations we use open boundary conditions, with 
$A$ sites at the left and the right extremes of a chain with $N_c$ unit cells.
}
\label{fig:fig1}
\end{figure}

The Hamiltonian of the system is schematically presented in Fig. \ref{fig:fig1} 
and is given by:
\begin{eqnarray}
 H&=&\sum_l\mathbf{A}_l\cdot(\mathbf{T}_l+\mathbf{T}_{l+1})+J\sum_l\left[ \frac{(T_l^2-s_l^2)}{4}\right.\nonumber\\
 & &+\left .\frac{1}{2}(\mathbf{T}_l\cdot\mathbf{T}_{l+1}+\mathbf{s}_l\cdot\mathbf{s}_{l+1})\right]-hS^z,
 \label{eq:ham-spin}
\end{eqnarray}
where $h$ is an applied magnetic field in the $z$-direction, with $g\mu_B\equiv1$, $l$ indexes unit cells, $S^z$ is the $z$ component of the total spin,
$\mathbf{A}_l$ is a spin-1/2 operator at the sublattice $A$ at cell $l$, and
we use the symmetrical and the anti-symmetrical composition of the spin-1/2 at $B$ sites in the same unit cell: $\mathbf{T}_l=\mathbf{B}_{1,l}+\mathbf{B}_{2,l}$ and
$\mathbf{s}_l=\mathbf{B}_{1,l}-\mathbf{B}_{2,l}$, respectively.
This Hamiltonian is in the class of diamond chain models \cite{refe1,refe2,refe3,refe4}. These were used to understand the interesting low-temperature properties of the mineral Azurite \cite{refe5,refe6,refe7,refe8,refe9}
and compounds of formula A$_3$Cu$_3$AlO$_2$(SO$_4$)$_4$ (with A = K, Rb, Cs) \cite{refe10,refe11,refe12}. The superexchange coupling between $A$ and $B$ spins is taken as the energy unit in this investigation, while $J$ is the coupling
between $B$ spins. The $A-B$ couplings have an $AB_2$ pattern \cite{Coutinho-Filho2008}, favoring a ferrimagnetic ordering for $h=0$, and a triplet state between 
$B$ sites at the same unit cell. The $J$ couplings, with a ladder arrangement \cite{Dagotto1996,PhysRevB.73.214405}, introduce frustration 
in the system since they favor a singlet pairing between $B$ sites at the same unit cell for $h=0$.

We use DMRG to characterize the zero temperature state of chains with $N_c$ unit cells and open boundary conditions, $A$ sites at
the two extremes.
We kept from 364 to 500 states per block in the DMRG calculations, and the typical discarded weight was $10^{-10}$, with a maximum of $10^{-8}$.

\section{The first-order phase transition and its energy curves}
\begin{figure}[!htb]
\includegraphics[width=0.47\textwidth]{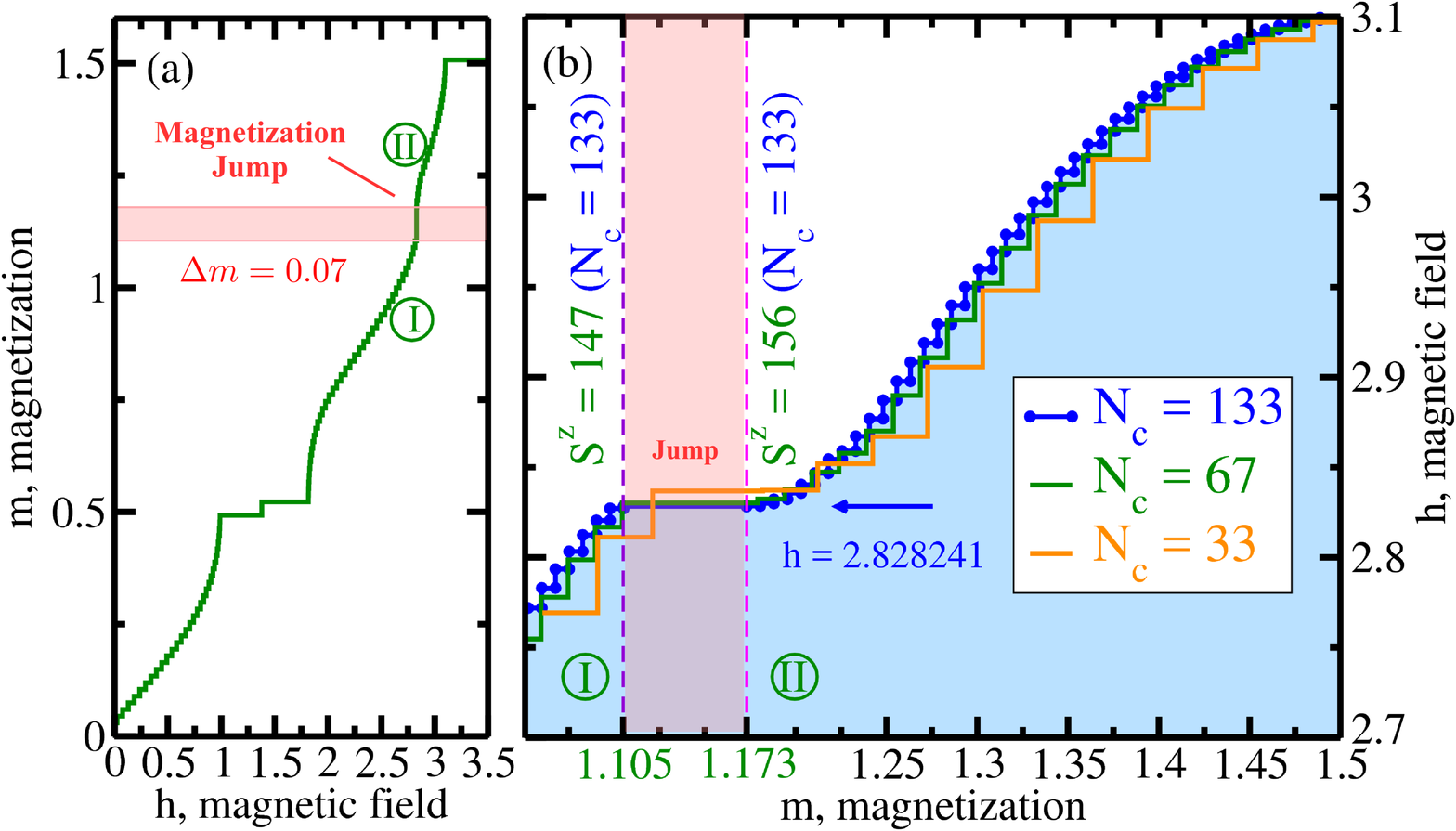}
\includegraphics[width=0.47\textwidth]{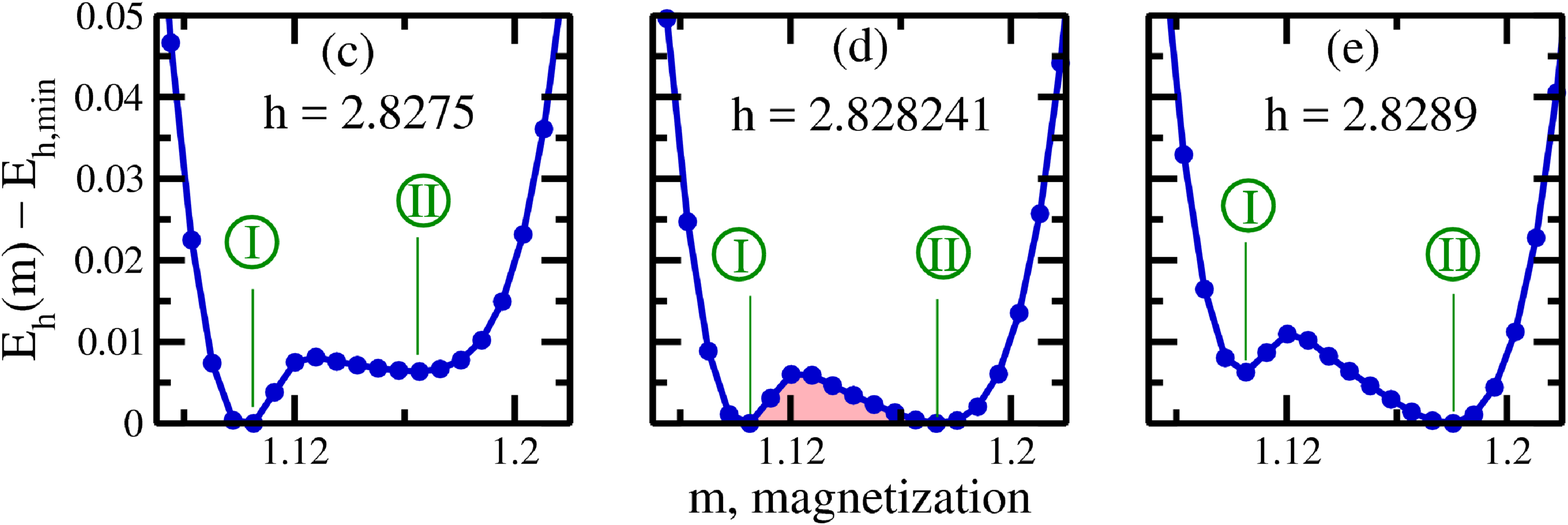}
\caption{First-order transition induced by a magnetic field $h$, $J=0.7$. 
(a) Magnetization per unit cell $m$ as a function of $h$ for 
a chain with $N_c=67$. The finite size implies a 
minimum magnetization step $\Delta m_{min}=1/N_c=1/67=0.015$. A jump of size $\Delta m=0.07$ at $B=2.83$
separates two competing quantum phases, I and II.
(b) $h$ as a function of $m$ for $N_c=33,~67,~133$ in the vicinity 
of the jump. For $N_c=133$, the jump occurs
between $S^z=147$ ($m=1.105$) and $S^z=156$ ($m=1.173$), where $S^z$ is the $z$ component of the total spin.
(c-e) $h$ fixed: the difference between the total energy $E_h(m)$ for a given $m$  
and the lowest energy value $E_{h,min}$, with $N_c=133$. A single global minimum is observed for $h$ 
(c) below ($h=2.8275$, phase I) or (e) above ($h=2.8289$, phase II) the jump. 
(d) There are two global minima for $h$ at the jump ($h=2.828241$). 
We kept a maximum of 364 states per block in the calculations.
}
\label{fig:fig2}
\end{figure}

We present in Fig. \ref{fig:fig2}(a) the magnetization per unit cell $m(h)$ as a function of $h$ for $J=0.7$ and $N_c=67$.
A magnetization jump of size $\Delta m=0.07$ is observed for $h=2.83$; in fact,
due to the finite size of the system, the $m(h)$ curve is made of finite size steps of size $1/N_c=1/67=0.015$, lower than $\Delta m$. 
This jump reveals the first-order transition between two competing phases, 
which we identify as I, before the jump, and II, after the jump, for increasing values of $h$.
In Fig. \ref{fig:fig2}(b) we show that this jump is robust to the thermodynamic limit by comparing the $h(m)$ curves for 
$N_c=33,67,\text{ and }133$. For $N_c=133$ the jump occurs between states with $S^z=147$ ($m=1.105$) and $S^z=156$ ($m=1.173$) at $h=2.828241\equiv h_t(J)$, the transition field for $J=0.7$.

We present the curves of energy as a function of $m$, $E_h(m)$, in Figs. \ref{fig:fig2}(c-e) for $N_c=133$. 
The $E_h(m)$ curve has a single global minimum for a value of $h$ lower or higher than $h_t$, 
as shown in Figs. \ref{fig:fig2} (c) and (e) for $h=2.8275$ and $h=2.8289$, respectively.
These are homogeneous phases with total spin $S^z<147$ (phase I), for $h\lesssim h_t$, and $S^z>156$, for $h\gtrsim h_t$ (phase II).
For $h=h_t$, Fig. \ref{fig:fig2}(d), the $E_h(m)$ curve shows two global minimum at $S^z=147$ ($m_I=1.105$) and $S^z=156$ ($m_{II}=1.173$). 
The states between these two values are metastable, or unstable with a negative compressibility $\partial^2 E_h(m)/\partial m^2$. 
The magnetization jump thus occurs between these two magnetization values and has a size $\Delta m=m_{II}-m_{I}=0.07$.
Since $E_h(m_I)=E_h(m_{II})$, the thermodynamic equilibrium state for $m_{I}<m<m_{II}$ is a phase-separated state 
with phases I and II coexisting in distinct spaces of the chain, and the thermodynamically stable $E_h(m)$ curve is 
flat between the two minima (double tangent method). We observe that if the $m(h)$ curve is built from the local 
relation $h=\partial E_{h=0}/\partial m$, a van der Waals loop curve will be obtained and the jump is 
determined after the use of Maxwell construction \cite{Kohno1997,Sakai1999}.

\section{Competing states and phase separation} 
\begin{figure}[!htb]
\includegraphics[width=0.4\textwidth]{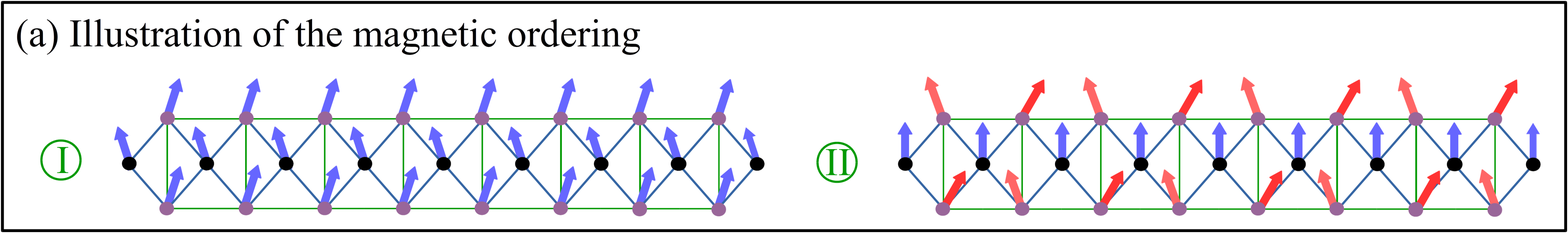}
\includegraphics[width=0.47\textwidth]{fig3bg.eps}
\includegraphics[width=0.4\textwidth]{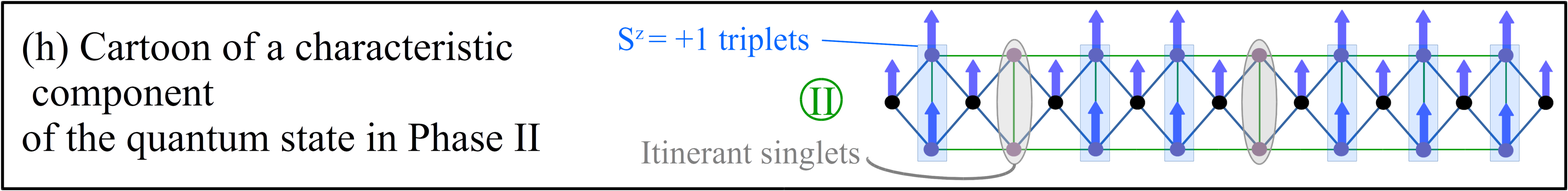}
\caption{The two competing states in the vicinity of the jump, $N_c=133$ and $J=0.7$. 
(a) Illustration of the magnetic orientation of $A$ and $B$ sublattices in phases I and II, as 
suggested by the data in (b-g). (b and c) Average spin of (c) $A$ and (d) $B_{1}+B_{2}$ 
sites at cell $l$, $\expval{A^z_l}$ and $\expval{T^z_l}=\expval{B_{1,l}^z}+\expval{B_{1,l}^z}$, respectively, 
in phases I ($S^z=140$) and II ($S^z=164$). 
(d and e) Transverse correlation function $C_T(r)$ as a function of the distance $r$ between cells: 
(d) in phase I there is a critical ferromagnetic correlation between $A$ spins and between $B_1$ spins (and $B_2$ spins, 
by symmetry), and an alternating critical correlation between $A$ and $B_1$ (and $B_2$) spins (inset);
(e) in phase II, $A$ spins are fully polarized and uncorrelated, while $B_1$ (and $B_2$) spins have an alternating 
critical transverse correlation (inset).
(f) Average correlation function between $B$ 
spins at the same unit cell $l$, $\expval{\mathbf{B}_{1,l}\cdot\mathbf{B}_{2,l}}$: in phase I, $B$ 
spins at the same unit cell are approximately aligned, while in phase II they are canted. 
(g) The average local singlet density $\eta_l=0.25-\expval{\mathbf{B}_{1,l}\cdot\mathbf{B}_{2,l}}$ evidence
the increase in the number of itinerant singlet pairs between $B$ spins at the same unit cell from phase I to phase II.
(h) Cartoon of a characteristic component of the quantum state in phase II: 
a box (ellipse) identifies a triplet state $\ket{\uparrow\uparrow}_l$ (a singlet 
state $(\ket{\downarrow\uparrow}_l-\ket{\uparrow\downarrow}_l)/\sqrt{2}$) between $B_{1,l}$ and $B_{2,l}$ 
spins at the cell $l$.
We kept a maximum of 364 states per block in the calculations.
}
\label{fig:fig3}
\end{figure}

In Fig. \ref{fig:fig3} (a) we illustrate 
the magnetic orientation of phases I and II, as suggested by the data shown in Figs. \ref{fig:fig3}(b)-(g), to 
help in the discussion.
The magnetization profile of $A$ spins $\expval{A^z_l}$ is homogeneous in the two phases if we discard sites near the 
boundaries, with $A$ spins approximately fully polarized in phase II, as shown in Fig. \ref{fig:fig3}(b). The magnetization profile of the $B$ 
spins is presented in Fig. \ref{fig:fig3}(c) through the sum 
\begin{equation}
\expval{T^z_l}=\expval{B^z_{1,l}}+\expval{B^z_{2,l}}.
\end{equation}
Considering the bulk, the profiles 
are also homogeneous, but the average magnetization of $B$ spins \textit{decreases} from phase I to phase II. Also, the order parameters of 
$A$ and $B$ sublattices become uncoupled after the transition. 

We also calculate the transverse spin correlation functions 
\begin{equation}
C_T(r)=\expval{\expval{S^x_m S^x_{n}+S^y_m S^y_{n}}}, 
\end{equation} 
with $|m-n|=r$, along the $A$ sublattice and along the $B_1$ sublattice, averaging 
the correlations among all pair of cells $m$ and $n$ such that $|m-n|=r$. In phase I, Fig. \ref{fig:fig3}(d), $C_T(r)$ shows a behavior consistent with a 
uniform ferromagnetic critical power-law behavior along each sublattice; while the correlation between spins in one sublattice with spins in the other, inset 
of Fig. \ref{fig:fig3}(d), is critical and alternating. These results show that in phase I the spin-flop transition occurs with 
a canting orientation between $A$ and $B$ spins, as illustrated 
in Fig. \ref{fig:fig3}(a). 
In phase II, Fig. \ref{fig:fig3}(e), the transverse correlation between 
$A$ spins is negligible, while $B_1$ spins have an alternating critical correlation, as shown 
in the inset of Fig. \ref{fig:fig3}(e). So, in phase II, the spin-flop transition occurs by a canting orientation between $B$ spins 
at the same sublattice, as illustrated in Fig. \ref{fig:fig3}(a).
\begin{figure}
\includegraphics*[width=0.47\textwidth]{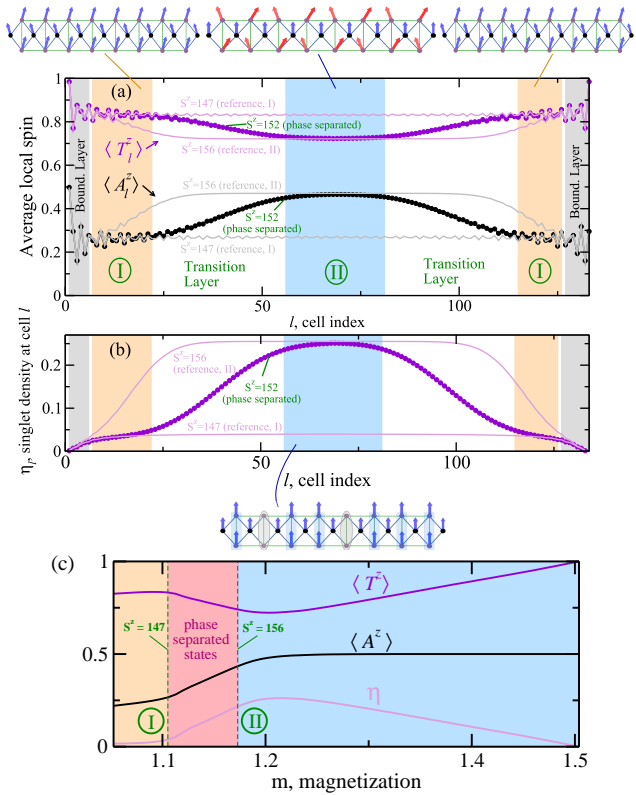}
\includegraphics*[width=0.4\textwidth]{fig4c.eps}
\caption{(a) and (b). Local averages in a \textit{phase-separated} state, $S^z=152$, $N_c=133$ and $J=0.7$. 
We present the data for $S^z=147$ and $S^z=156$ as a reference.
The open boundaries of the chain dominate the averages in  
\textit{boundary layers}, while \textit{transition layers} exhibit a mixed phase.
We present the schematic magnetic orientation in the two coexisting regions 
and a cartoon of the quantum state in phase II.
(a) Average local spin of $A$ sites, $\expval{A^z_l}$, and $B_{1,l}+B_{2,l}$ sites, 
$\expval{T^z_l}=\expval{B_{1,l}^z}+\expval{B_{1,l}^z}$, at cell $l$.
(b) Average singlet density $\eta_l=0.25-\expval{\mathbf{B}_{1,l}\cdot\mathbf{B}_{2,l}}$ at cell $l$. 
To stabilize the state with $S^z=152$,
the calculation required 48 sweeps, keeping 500 states in the last 6. (c) Global averages as a function 
of magnetization, $N_c=133$ and $J=0.7$: $\expval{T^z}=\frac{1}{L}\sum_{l}\expval{T^z_l}$, $\expval{A^z}=\frac{1}{L}\sum_{l}\expval{A^z_l}$, and $\expval{\eta}=\frac{1}{L}\sum_{l}\expval{\eta_l}$, where $l=[8,125]$ and $L=118$. We kept a maximum of 364 states per block in the DMRG calculations shown in (c).
}
\label{fig:fig4}
\end{figure}

The correlation between 
$B$ spins at the same cell, $\expval{\mathbf{B}_{1,l}\cdot\mathbf{B}_{2,l}}$, is shown in Fig. \ref{fig:fig3}(f) and evidence that in phase I these spins are in a superposition of triplet states: $\ket{\downarrow\downarrow}_l$, 
$(\ket{\downarrow\uparrow}_l+\ket{\uparrow\downarrow}_l)/\sqrt{2}$, and $\ket{\uparrow\uparrow}_l$, since $\expval{\mathbf{B}_{1,l}\cdot\mathbf{B}_{2,l}}=0.25$ for these states. In particular, the component $\ket{\uparrow\uparrow}$ is the most relevant 
because $\expval{T^z_l}\approx 0.8$, as shown in Fig. \ref{fig:fig3}(c).
In phase II, $\expval{\mathbf{B}_{1,l}\cdot\mathbf{B}_{2,l}}\neq 0.25$ and $B_1$ and $B_2$ spins at the same cell are canted, one related to the other, 
as illustrated in Fig. \ref{fig:fig3}(a). To figure out an approximate 
picture of the quantum state, we remember that the four quantum states 
of the pair are the three triplet states and the singlet state, $(\ket{\downarrow\uparrow}_l-\ket{\uparrow\downarrow}_l)/\sqrt{2}$, 
with $\expval{\mathbf{B}_{1,l}\cdot\mathbf{B}_{2,l}}= -0.75$. Thus we expect that the quantum state in phase II has a large number of components in 
which the $B$ spins at the same cell are in a singlet state. 

We define the average local singlet density as 
\begin{equation}
\eta_l\equiv 0.25-\expval{\mathbf{B}_{1,l}\cdot\mathbf{B}_{2,l}}
\end{equation}
and present it in Fig. \ref{fig:fig3}(g). In phase II, $\eta_l$ has an average value of 0.25, one singlet spin pair for every four cells, and
is homogeneous throughout the chain; while in phase I, also homogeneous, $\eta_l$ has a much lower value. We illustrate the quantum state in phase II 
in Fig. \ref{fig:fig3}(h).
In fact, we attribute the oscillations in the magnetizations shown in Figs. \ref{fig:fig3}(b) and (c) to the itinerancy of these singlet pairs 
throughout the chain, with a low singlet density in phase I and a high singlet density in phase II. 

In Fig. \ref{fig:fig4} we present the local properties of the low energy state in the $S^z=152$ sector, which has a value of 
magnetization $m=1.143$ between $m_I$ and $m_{II}$. The magnetization profiles shown
in Fig. \ref{fig:fig4}(a) are compared with the ones in states just before the jump, with $S^z=147$ ($m=m_I$), and just after the jump, with $S^z=156$ ($m=m_{II}$). 
In the profile of $S^z=152$, we identify the magnetizations of phases I and II in spatially separated regions. The \textit{transition layers} between the two phases, 
the interface between then, have a mixed state with the coexistence 
of the two phases in the same spatial region and is relatively large. Further, \textit{boundary layers} are spatial regions strongly dominated by the boundaries 
of the system, in which the magnetizations do not change appreciably among the three states. In Fig. \ref{fig:fig4}(b) 
we show that the singlet density $\eta_l$ enforce these conclusions. In fact, the singlet density profile at the center of the chain in the 
state with $S^z=152$ coincides with that of phase II ($S^z=156$), while the profile changes to that of phase I after the transition layers. 

In Fig. \ref{fig:fig4}(c), we show the global average of the magnetizations of $A$ and of $B$ sites, as well as that of the 
singlet density:
\begin{equation}
 \expval{A^z}=\frac{1}{L}\sum_{l}\expval{A^z_l}
\end{equation}

\begin{equation}
 \expval{T^z}=\frac{1}{L}\sum_{l}\expval{T^z_l},
\end{equation}

\begin{equation}
\expval{\eta}=\frac{1}{L}\sum_{l}\expval{\eta_l},
\end{equation}
respectively, as a function of the magnetization from below the jump up to the saturation magnetization. 
To discard the cells in the boundary layers, we consider $l=[8,125]$, such that $L=118$. 
The behavior of these averages is approximately linear for magnetizations inside the jump. 
However, this behavior depends on the range of sites chosen to calculate these quantities due to 
the inhomogeneous nature of the states inside the jump.
Further, we observe that the average singlet density monotonically decreases with $h$ after the transition field, with the magnetization of $B$ spins increasing in accord, and the $A$ spins remaining fully polarized up to the saturation field.
Using the singlet density, we can make 
a direct comparison of this transition with a thermal liquid-gas transition by identifying the low-density phase, phase I, as the gas phase and the 
high-density phase, phase II, as the liquid phase. 

We attribute the origin of the transition to the competition between the $A-B$ superexchange pattern and 
$J$ couplings, ladder superexchange pattern, in the Hamiltonian (\ref{eq:ham-spin}), and the magnetic field. 
In phase I, the magnetization of $A$ spins is less than 1/2 due to the antiferromagnetic coupling with $B$ spins, which have the triplet 
$\ket{\uparrow\uparrow}$ as the most relevant component. Notice that 
The $J$ coupling between these spins does not favor this component. 
In phase II, $A$ spins are fully polarized, thanks to the magnetic field, and the dynamics of the system is governed by the ladder $J$ couplings, with a larger singlet density. The energy of these two phases becomes equal at the transition field, with a discontinuity 
in the first derivative at the transition. The order of the transition and phase separation can be better understood by considering the dynamics of the singlets, which act like holes in a doped system. The ladder pattern 
in Hamiltonian (\ref{eq:ham-spin}) has terms \cite{PhysRevB.73.214405} related to the itinerancy of the singlets, which 
are hard-core bosons, and a repulsion between singlets in nearest neighbor cells. Consider a fixed value of singlet density, $\expval{\eta}$, between
its value in phase I, $\expval{\eta}_{I}$, and in phase II, $\expval{\eta}_{II}$, for which the phase-separated state has lower energy. 
If $\expval{\eta}\gtrsim \expval{\eta}_I$, a nucleation process of phase I takes place by the expelling of some singlets from a region of the chain, 
lowering the singlet density in this region, thereby increasing the density in the complementary region of the chain. This process can be understood as an effective attraction between the singlets. This transfer of singlets stops when the first portion of the chain has a density $\expval{\eta}_I$ and the second portion $\expval{\eta}_{II}$. The specific 
value of $\expval{\eta}_{II}$ is due to the hard-core constraint and the repulsion between the singlets in neighboring cells.  
\begin{figure}
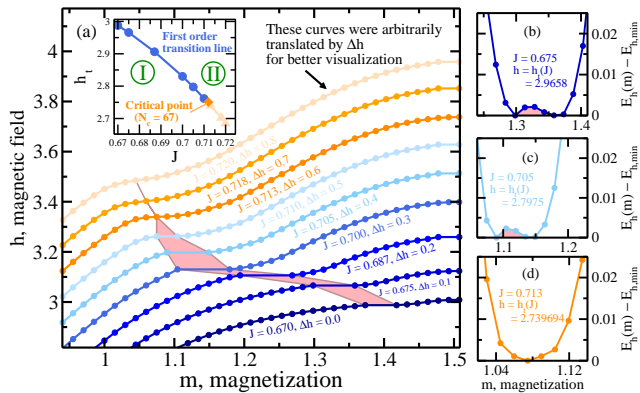

\includegraphics*[width=0.34\textwidth]{fig5a.eps}
\includegraphics*[width=0.122\textwidth]{fig5bd.eps}
\caption{Closing of the magnetization jump and the critical point, $N_c=67$.
(a) $h$ as a function of $m$. We present the central value of $h$ in the magnetization 
finite-size plateaus, except for the plateaus bounding the jump. We applied an arbitrary translation $\Delta h$ 
to $h$ for a better visualization. The magnetization jump decreases to the 
minimum attainable value for $N_c=67$, $\Delta m_{min}=0.015$, at $J\approx0.712\equiv J_c^{(N_c)}$. In the Inset we present the transition line $h_t(J)$ with
first-order transition points, the critical point, and the crossover line (triangles). (b-d) Difference between the total energy $E_h(m)$ for a fixed $h$ 
and its lowest value $E_{h,min}$ for the same $h$ and $J$. The curves are shown 
for (b) $J=0.675$, (c) $J=0.705$, and (d) $J=0.713$, with the respective $h=h_t(J)$. The double minima observed for $J$ below
the critical value $J<J_{c}^{(N_c)}$, 
as in (b) and (c), change to a single minimum for $J>J_{c}^{(N_c)}$, as shown in (d). We kept a maximum of 364 states per block in the calculations.}  
\label{fig:fig5}
\end{figure}

At a critical value of $J$, a continuous second-order transition occurs between the two phases. 
In Fig. \ref{fig:fig5}(a) we present $h(m)$ curves for values of $J$ near $J=0.7$ and $N_c=67$, with an arbitrary 
translation in $h$ at each curve for a better visualization. The magnetization jump exhibits a little increase as $J$ increases from $J=0.675$ 
to $J=0.700$, and starts to decrease at $J=0.705$. The jump closes between $J=0.710$ and $J=0.713$, implying a critical 
point at $J_c^{(N_c=67)}=0.712$; since energies curves with a double minima, Figs. \ref{fig:fig5}(b) and (c), changes to a single minimum curve in 
the critical point, Fig. \ref{fig:fig5}(d). In the inset of Fig. \ref{fig:fig5}(a) we draw the $h-J$ phase diagram. The first-order transitions 
are defined by the value of $h$ at which the jump occurs, for which $(\partial h/\partial m)=0$. At the critical point, the same 
condition $(\partial h/\partial m)=0$ applies; while for $J>J_c^{(N_c=67)}$ the crossover line between the two phases 
is defined by the minimum in the $(\partial h/\partial m)$ curve.   

\section{Summary} 
We studied the first-order quantum transition induced by
a magnetic field in a quantum frustrated ferrimagnetic chain through the density matrix renormalization group. 
The first-order transition gives rise to a magnetization jump in the magnetization curve of the system. 
In particular, we did a detailed analysis 
of the phase-separated states inside the jump for a frustration parameter $J=0.7$.
The Hamiltonian of the system has three spins per unit cell and is in the class of 
diamond chain or frustrated $AB_2$ chain models.
The energy curves as a function of the magnetization present a double global minimum in the transition field, 
with unstable and metastable states. The first-order transition occurs between states with
different densities of itinerant singlet paired spins, from a state with a lower
singlet density (``gas'' phase) to a higher one (``liquid phase''), as evidenced by the local 
averages of the magnetization, local correlations and transverse correlation functions along the chain.
These quantities show that in the ``liquid phase'', two spins in the unit cell are canted, 
and present evidence of a power-law alternating transverse correlation along the chain, while the third spin is approximately 
fully polarized. Local averages of the magnetization and the singlet density show that the states inside the jump are phase separated with the two phases observed in distinct spatial regions. 
The interface between the two spatially separated phases is in a mixed phase with the two phases
coexisting in the same spatial region. Further, the global averages of the same quantities have an approximately 
linear behavior with the magnetization inside the jump. In particular, the singlet density increases with magnetization, 
while the magnetization of the third site increases. For higher magnetizations, above the jump, the singlet density decreases 
monotonically to zero up to the saturation magnetization. 
In the $h-J$ phase diagram, the first-order transition line ends at a critical point beyond
which a crossover between the two phases takes place. The competition between the superexchange couplings, frustration, 
is an essential ingredient to the first- and second-order transitions.  

Some interesting questions can be addressed by future research on this model. 
In particular, we mention the study of the hysteresis loops induced by the metastable states, 
including their dynamical and thermodynamical aspects, as well as a detailed analysis of the critical region.   

We acknowledge support from Coordena\c{c}\~ao de Aperfei\c{c}oamento de Pessoal de N\'{\i}vel Superior (CAPES),
Conselho Nacional de Desenvolvimento Cient\'{\i}fico e Tecnol\'ogico (CNPq), and Funda\c{c}\~ao de Amparo \`a Ci\^encia e
Tecnologia do Estado de Pernambuco (FACEPE), Brazilian agencies, including the PRONEX Program which is funded by
CNPq and FACEPE, APQ-0602-1.05/14.

%\bibliography{bibliografia}
%

\end{document}